\documentstyle[12pt]{article}

\begin{document}
%\baselineskip=1 truecm
\def\bib#1{[{\ref{#1}}]}
\def\at{\tilde{a}}

\begin{titlepage}
 \title{ Quantum Groups and Quantum Field Theory\\ 
                  in Rindler Space-Time }

\author{Gaetano  Lambiase \thanks{E-mail: lambiase@vaxsa.csied.unisa.it}\\ 
{\em  Dipartimento di Fisica Teorica e S.M.S.A.}\\ 
{\em  Universit\`a di Salerno, 84081 Baronissi (Salerno), Italy.}\\
{\em  Istituto Nazionale di Fisica Nucleare, Sezione di Napoli.}}
              \date{\empty}
              \maketitle

              \begin{abstract}

Quantum Field Theory (QFT) developed in 
Rindler space-time 
and its thermal properties are analyzed by means of 
quantum groups approach.  
The quantum deformation parameter, labelling the unitarily 
inequivalent
representations, turns out to be related to the 
acceleration of the Rindler frame.  
              \end{abstract}

\thispagestyle{empty}
\vspace{20. mm}
PACS number(s): 02.20.-a, 04.62.+v. 
              \vfill
	      \end{titlepage}
\section{\bf Introduction}

In recent years there have been remarkable progresses and growing 
interest in two fields  of research: quantum gravity and
quantum groups (q-groups).

Although many attempts have been made to quantize gravity, a
satisfactory 
and definitive theory still does not exist. As well known indeed,
one of the most discussed problems is the 
non-renormalizability of General Relativity (GR),
or its various generalizations, when quantized as a local 
quantum field theory.
In the absence of a theory of quantum gravity, one can try to analyze 
quantum aspects of gravity by studying QFT in curved space-time, namely
by studying the quantization of matter fields in the presence of
the gravitational field as a classical 
background described by GR. 
 
An important result in this approach has been the discovery by
Hawking  
that quantum effects can lead to thermal evaporation of black holes
\cite{HAW}. Owing to this result, different 
background space-times have been investigated, putting special attention 
to the Rindler space-time, associated with an uniformly accelerated 
observer in Minkowski space-time. Davies \cite{DAV} and 
Unruh \cite{UNR} have
shown that the vacuum state for an inertial observer is a canonical
ensemble 
for an uniformly accelerated observer (Rindler observer). 
The temperature characterizing  this ensemble is related to the
acceleration of the observer by the relation
\begin{equation}
T\,=\,\frac{a}{2\pi}
\end{equation}
in units $\hbar=c=k_B=1$. This result  is known as {\it thermalization
theorem} 
(for a review, see \cite{TAK}). We note that the replacement of the 
acceleration with the surface of gravity of a black hole leads to the
Bekenstein-Hawking temperature.

The purpose of this paper is to show the formal connection between 
thermalization theorem and q-groups. Such a connection will be
established by means of the generator of Bogoliubov transformations
expressed in terms of quantum deformation operators 
of the Weyl-Heisenberg (q-WH) algebra.

One feature of physical interest which emerges from our analysis is that the
quantum deformation parameter is related to the Rindler acceleration;
hence, for the equivalence principle, to the static gravitational field.
From this result follows that 
the quantum deformation can be induced by the gravitational
background. Such a conclusion, although in a different context, has been
also observed in Ref. \cite{DIS}. 

Besides, as we will show, the 
Rindler acceleration labels the unitarily inequivalent representations
of the canonical commutation relations.
The existence of
the unitarily inequivalent representations plays a crucial
role in  
the quantization of matter fields in curved space-time \cite{SOD}
and in our case, in 
the quantization of a free scalar field in the Rindler manifold.

{}Here we will not present and discuss the properties of q-groups, 
being well studied in the literature. We recall 
only that q-groups are examples of quasi-triangular 
Hopft algebra \cite{HOP}. More specifically, q-groups are the
deformation of the universal enveloping algebra of a finite-dimensional
semi-simple Lie algebra. 
Due to its richer structure than that of Lie groups, q-groups provide 
a powerful mathematical tool in
different topics of modern physics: quantum optics \cite{OPT}, 
quantum dissipation \cite{DIS}, gauge field theory \cite{GFT}, 
quantum gravity \cite{KEM, MAG}, etc.

The layout of this paper is the following. In Section 2 we review the 
salient points of the thermalization theorem in order to make more 
transparent its connection with q-groups. 
In Section 3 we present, for one degree of freedom, a realization of 
the quantum deformation
of Weyl-Heisenberg algebra in term of finite difference
operator over the set of entire analytical functions
\cite{MAR}. 
Finally, in Section 4, we generalize the results of Section 3 to the
case of 
infinite degrees of freedom in order to establish the formal relation
between thermalization 
theorem and q-groups. Section 5 is devoted to the 
conclusions.

In this paper we do not study the role of the coproduct operation, nor 
do we investigate the superalgebra features of q-WH algebra in
connection 
with QFT in Rindler space-time. Such an analysis needs further and deeper 
formal investigation, which goes beyond the task of present paper
(concerned mainly to displaying the relation between the Rindler acceleration
and the q-deformation 
parameter) and which we plan 
for future work.

\section{Thermalization Theorem}
\setcounter{equation}{0}

In this Section, we shall briefly discuss the derivation of the relation 
between the inertial and accelerated description of free quantum fields
in 
flat space-time. We shall treat the case of a complex massive scalar 
quantum field $\phi(x)$ in n-dimensional Minkowski space-time. 

In the accelerated frame it is customary to use the Rindler coordinates
$(\eta, \xi,; \vec{x}_R)$ which are related to 
Minkowski coordinates $(x^0, x^1; \vec{x}_M)$ by transformations
\begin{equation}
x^0=\xi\sinh\eta,\,\,\, x^1=\xi\cosh\eta,\,\,\, \vec{x}_M=\vec{x}_R=
(x^2,\dots,x^{n-1}).
\end{equation}

The world line of a uniformly accelerated observer is described by
$\xi=const$. 
Its acceleration is $(\ddot{x}^{\mu}\ddot{x_{\mu}})^{1/2}=\xi=a^{-1}$
and 
its proper time $\tau$ is proportional to the Rindler time $\eta$, 
$\eta=a\tau$. The Rindler coordinates cover only two regions of the 
Minkowski
space: the Rindler wedge $R_{+}=\{x|x^1>|x^0|\}$
and 
the wedge $R_{-}=\{x|x^1<-|x^0|\}$.
 
The field $\phi(x)$ is solution of Klein-Gordon equation
$(\Box\,+\,m^2)\phi(x)=0$,
where
$\Box=(-g)^{-1/2}\partial_{\mu}(\sqrt{-g}g^{\mu\nu}\partial_{\nu})$.  

In the Minkowski space, the quantized field $\phi(x)$ can be
decomposed 
in Minkowski modes $\{U_k(x)\}$ ($k=(k_1, \vec{k})$) 
\begin{equation}
\phi(x)=\int d^{n-1}k\,[a_{k}U_k(x)\,+\,\bar{a}_k^{\dagger}U_k^{*}(x)]\,{.} 
\end{equation}
The set $\{U_k\}$ forms (with respect to the Klein-Gordon inner product) an
orthonormal basis; $a_k$ and $\bar{a}_k^{\dagger}$ operators satisfy the
canonical  
commutation relations
\begin{equation}
[a_k,
a_{k'}^{\dagger}]=[\bar{a}_k,\bar{a}_{k'}^{\dagger}]=\delta(k_1-k_1') 
\delta(\vec{k}-\vec{k}').
\end{equation}
All other commutators are zero.

The Hamiltonian operator reads $H_M=\int
d^{n-1}k\,\omega_k\, 
(a_k^{\dagger}a_k\,+\,\bar{a}_k\bar{a}_k^{\dagger})$, where $\omega_k=
\sqrt{k_1^2+|\vec{k}|^2+m^2}$.
The operators $a_k, \bar{a}_k$ and $a_k^{\dagger},\bar{a}_k^{\dagger}$ are
interpreted,  
respectively, as annihilation and creation operators for the states of
the  
Fock space constructed from the Hilbert space of Minkowski modes.
The Minkowski vacuum, denoted $|0_M>$, is defined by
\begin{equation}
a_k|0_M>\,=\,\bar{a}_k|0_M>\,=0, \quad \forall \,k\,{.} 
\end{equation}
In the Rindler coordinates, the quantization of the scalar field follows  
the same prescription of the Minkowski case \cite{FUL}. The quantum
field is  
expanded in terms of Rindler 
modes $\{u_k^{(\sigma)}(x), \sigma=\pm\}$ ($k=(\Omega,\vec{k})$)
\begin{equation}
\phi(x)=\int_{0}^{\infty}d\Omega\int d^{n-2}k\sum_{\sigma}\,[
b_k^{(\sigma)}u_k^{(\sigma)}(x)\,+\,\bar{b}_k^{(\sigma)\dagger}
u_k^{(\sigma)*}(x)],
\end{equation}
The set 
$\{u_k^{(\sigma)}\}$ is an orthonormal basis in the wedge $R_{\sigma}$.
The canonical commutation relations are
\begin{equation}
[b_k^{(\sigma)}, b_{k'}^{(\sigma')\dagger}]=
[\bar{b}_k^{(\sigma)}, \bar{b}_{k'}^{(\sigma')\dagger}]=
\delta_{\sigma\sigma'}\delta(\Omega-\Omega')\delta(\vec{k}-
\vec{k}')\,{.}
\end{equation}
All other commutators are zero. The Hamiltonian operator 
is $H_R\,=\,H_R^{(+)}\,-\,H_R^{(-)}$, with
\begin{equation}
H_R^{(\sigma)}=\int_0^{\infty} d\Omega\,\int d^{n-2}k\,\Omega\left[
b_k^{(\sigma)\dagger}
b_k^{(\sigma)}\,+\,\bar{b}_k^{(\sigma)}
\bar{b}_k^{(\sigma)\dagger}\right].
\end{equation}
In analogy to Minkowski case, $b_k^{(\sigma)}$,
$\bar{b}_k^{(\sigma)}$ 
and $b_k^{(\sigma)\dagger}$, $\bar{b}_k^{(\sigma)\dagger}$ are
interpreted 
respectively, as annihilation and creation operators 
for the states of the Fock space 
constructed from Hilbert space associated to Rindler modes. 
The Rindler vacuum, $|0_R>=|0_{(+)}>\otimes|0_{(-)}>$, is defined by
\begin{equation}
b_k^{(\sigma)}|0_R>\,=\,\bar{b}_k^{(\sigma)}|0_R>\,=\,0, \quad \forall
\sigma, k\,{.}
\end{equation}
Note that the proper energy of Rindler particles seen by an accelerated
observer is not $\Omega$, but $a\Omega=\tilde{\omega}$, because Rindler
modes depend on time as $e^{-i\Omega\eta}=e^{-i(a\Omega)\tau}$. Keeping
in mind this point and
equating the two expressions for the field $\phi(x)$, Eqs. (2.2) and
(2.5),  
one obtains the Bogoliubov transformations 
\begin{equation}
b_k^{(\sigma)}=\sqrt{1+N(\tilde{\omega}/a)}\,d_k^{\,(\sigma)}\,+\,
\sqrt{N(\tilde{\omega}/a)}\,
\bar{d}_{\tilde{k}}^{\,(-\sigma)\dagger}\,{,}
\end{equation}
\begin{equation}
\bar{b}_{\tilde{k}}^{(-\sigma)\dagger}=\sqrt{N(\tilde{\omega}/a)}\,
d_k^{\,(\sigma)}\,+
\,\sqrt{1+N(\tilde{\omega}/a)}\,
\bar{d}_{\tilde{k}}^{\,(-\sigma)\dagger}\,{,}
\end{equation}
where $k=(\tilde{\omega}, \vec{k})$,
$\tilde{k}=(\tilde{\omega},-\vec{k})$,  
$N(\tilde{\omega}/a)=
(e^{2\pi\tilde{\omega}/a}-1)^{-1}$ and $-\sigma=-(\pm)=\mp$. 
The operators $d$ and $\bar{d}$ are related to
Minkowski operators $a$ and $\bar{a}$ by the relation
\begin{equation}
d_{\tilde{\omega},\vec{k}}^{\,(\sigma)}=
\int_{-\infty}^{+\infty}dk_1\left[
\frac{1}{\sqrt{2\pi\omega_k}}\left(\frac{\omega_k+k_1}{\omega_k-k_1}
\right)^{i\sigma\tilde{\omega}/2}\right]\,a_{k_1,\vec{k}},
\end{equation}
and analogous expression for $\bar{d}^{\,(-\sigma)}_{\tilde{k}}$. 
These operators annihilate the Minkowski vacuum
\begin{equation}
d_k^{\,(\sigma)}|0_M>\,=
\,\bar{d}_{\tilde{k}}^{\,(-\sigma)}|0_M>\,=\,0,\quad
\forall\, \sigma, k,
\end{equation}
and satisfy the canonical commutation relations. Transformations (2.9)
and  
(2.10) will turn out to play a fundamental role in connection with
quantum-groups. 
Bogoliubov transformations and the {\it ansatz} $|0_M>=
F(b^{\dagger},\bar{b}^{\dagger})|0_R>$ (F is a function to be
determined) 
allow to relate the Minkowski vacuum to the Rindler vacuum
\begin{equation}
|0_M>=Z\,\exp\left[{\int_{0}^{\infty} d\tilde{\omega}\,
\int d^{n-2}k\sum_{\sigma}\,
e^{-\pi\tilde{\omega}/a}
b_{k}^{(\sigma)\dagger}\bar{b}_{\tilde{k}}^{(-\sigma)\dagger}}
\right]|0_R>\,{,}\end{equation}
where $Z$ is a normalization constant. The physical meaning of
Eq. (2.13) is  
that the Minkowski vacuum can be expressed as a coherent state of
Cooper-like  
pairs of quanta in the Rindler wedges $R_+$ and $R_-$. For an operator 
$O_R$ localized in the Rindler wedge one finds
$<0_M|O_R|0_M>\,=\,\mbox{tr}(\rho\,O_R)$,
where $\rho$ is the density matrix
$\rho=e^{-aH_R/T}/\mbox{tr}e^{-aH_R/T}$.
$H_R$ is the Rindler Hamiltonian and $T=a/2\pi$. From here the 
thermalization theorem follows.

Our purpose in this paper is to show that the generator of the Bogoliubov
transformations (2.9) and (2.10) can be expressed in terms of operators
of the q-WH algebra, thus establishing the link between thermalization
theorem and q-groups.

\section{q-Deformation of the W-H Algebra}
\setcounter{equation}{0}

In this Section we shall focus our attention on the main features of the 
q-WH algebra in terms of finite difference operator. 

The Weyl-Heisenberg algebra is generated by the set of operators 
$\{\alpha,\alpha^{\dagger},I\}$ 
\begin{equation}
[\alpha, \alpha^{\dagger}]=I,\quad [N, \alpha]=-\alpha,\quad 
[N, \alpha^{\dagger}]=
\alpha^{\dagger}\,{,}
\end{equation}
with $N=\alpha^{\dagger}\alpha$ and all other commutators equal to zero. 
The Fock space ${\cal H}$ is generated by operating with $\alpha^{\dagger}$
on the vacuum 
$|0>$ ($\alpha|0>=0$).
The q-WH algebra is generated by
operators 
$\{\alpha_q,\hat{\alpha}_q, N_q
\equiv N, 
q\in {\cal C}\}$ \cite{BRI}
\begin{equation}
[\alpha_q, \hat{\alpha}_q]=q^{N},\quad
[N, \alpha_q]=
-\alpha_q,\quad [N, \hat{\alpha}_q]=
\hat{\alpha}_q\,{.}
\end{equation}
In the limit $q\to 1$, $\alpha_q \to \alpha, 
\hat{\alpha}_q \to \alpha^{\dagger}$. By following 
Refs. \cite{DIS, MAR}, 
it is convenient to work in the
space ${\cal F}$ of analytic functions because in ${\cal F}$ the
analytic  
properties of the Lie algebra structure are preserved under
q-deformation. To  
this end, we will adopt the Fock-Bargmann representation \cite{PER} in
which 
the operators
\begin{equation}
\alpha^{\dagger}\to z,\quad \alpha\to \frac{d}{dz},\quad N\to
z\frac{d}{dz},\,\,\, z\in {\cal C}\,{,}
\end{equation}
provide a realization of the WH algebra (3.1). In this context, the
space ${\cal F}$ becomes isomorphic to ${\cal H}$ and has a 
well defined inner 
product. A generic wave function $\psi(z)$ is expressed as
\begin{equation}
\psi(z)=\sum_{n}c_n u_n(z)\,{,} \qquad u_n(z)=\frac{z^n}{\sqrt{n!}}, 
\quad z\in{\cal C}, n\in {\cal N}. 
\end{equation}
The set $\{u_n(z)\}$ forms an orthonormal basis in ${\cal F}$.
The conjugation of the operators is defined with respect to the inner
product 
in ${\cal F}$. In order to provide a realization of the q-WH algebra in
the 
Fock-Bargmann representation, let us introduce the finite difference
operator 
${\cal D}_q$ (q-derivative operator) \cite{MAR, FIV}, defined as
\begin{equation}
{\cal D}_q
f(z)=\frac{f(qz)-f(z)}{(q-1)z}=\frac{q^{z\frac{d}{dz}}-1}{(q-1)z}f(z)\,{,} 
\end{equation}
where $f(z)\in {\cal F}$. ${\cal D}_q$ reduces to standard derivative in
the limit 
$q\to 1$. The set of operators $\{z, \frac{d}{dz}, {\cal D}_q\}$
satisfies the algebra
\begin{equation}
[{\cal D}_q, z]= q^{z\frac{d}{dz}},\quad [z\frac{d}{dz}, {\cal D}_q]=
-{\cal D}_q, \quad [z\frac{d}{dz}, z]=z.
\end{equation}
In terms of the operators $\{\alpha_q, \hat{\alpha}_q,
N_q\equiv N\}$, with
\begin{equation}
N\to z\frac{d}{dz},\quad \hat{\alpha}_q\to z,\quad 
\alpha_q\to{\cal D}_q,
\end{equation}
the finite difference operator algebra (3.6) becomes
\begin{equation}
[N, \alpha_q]=-\alpha_q,\quad [N, \hat{\alpha}_q]=
\hat{\alpha}_q,\quad 
[\alpha_q, \hat{\alpha}_q]=q^N\,{.}
\end{equation}
The algebra (3.8) thus provide a realization of the quantum WH algebra. 
Following Refs. \cite{DIS, MAR}, the commutator $[\alpha_q, 
\hat{\alpha}_q]=q^N$ can be formally 
written in ${\cal F}$ as 
\begin{equation}
[\alpha_q, \hat{\alpha}_q]=
\sqrt q\,e^{\frac{\epsilon}{2}
(\alpha^{2}-\alpha^{\dagger 2})}\equiv \sqrt q\,{\cal S}(\epsilon),
\quad q\equiv e^{\epsilon}, \epsilon\,\, real.
\end{equation}
${\cal S}(\epsilon)$ generates the following transformation 
\begin{equation}
\alpha
\to \alpha(\epsilon)={\cal S}(\epsilon)\alpha{\cal S}^{-1}(\epsilon)=
\alpha\cosh\epsilon+\alpha^{\dagger}\sinh\epsilon\,{,}
\end{equation}
and its hermitian conjugate. We finally observe that ${\cal
S}(\epsilon)$ is 
an element of the group $SU(1,1)$. In fact, by defining $J_{-}=\frac{1}{2}
\alpha^{2}, 
J_{+}=\frac{1}{2}\alpha^{\dagger\,2}, 
J_{0}=\frac{1}{2}(\alpha^{\dagger}
\alpha+\frac{1}{2})$, the $su(1,1)$ algebra is closed.
 
In the limit $\mbox{Im}\{z\}\to 0$ the above Fock-Bargmann representation 
scheme gives the Schr\"{o}dinger representation \cite{DIS}.

Eq. (3.9) is the key relation for establishing the link between thermalization
theorem and q-WH algebra.

\section{Thermalization Theorem by q-Groups}
\setcounter{equation}{0}

In this Section the connection between q-groups and thermalization
theorem 
will be established. Such a connection is obtained by extending the
q-WH 
algebra, discussed in Section 3, from one degree of freedom to infinite  
degrees of freedom. Therefore, some preliminary considerations will be
useful.

As pointed out in Section 2, Rindler coordinates cover two disconnected 
regions of the Minkowski space-time. Then for each Rindler wedge there
are 
two couples of annihilation and creation operators, one for particles
and one for anti-particles. 

This suggests 
to apply results of Section 3 to the set of four operators, 
$(\alpha^{(\sigma)}, \beta^{(\sigma)})$, $\sigma=\pm$, such that 
$\alpha^{(\sigma)}|0_M>=\beta^{(\sigma)}|0_M>=0$ and satisfying the
canonical  
commutation relations
\begin{equation}
[\alpha^{(\sigma)},\alpha^{(\sigma')\dagger}]=
[\beta^{(\sigma)}, \beta^{(\sigma')\dagger}]=\delta_{\sigma\sigma'}\quad
\sigma,\sigma'=\pm\,{.}
\end{equation}
Repeating for such operators the procedure shown in Section 3 leading to 
Eq. (3.9), one gets
\begin{equation}
[\alpha_q^{(\sigma)}, \hat{\alpha}_q^{(\sigma)}]=\sqrt{q}
e^{\frac{\epsilon}{2}(\alpha^{(\sigma)2}-\alpha^{(\sigma)\dagger\,2})}\equiv
\sqrt{q}\,{\cal S}_1(\epsilon),\quad q\equiv e^{\epsilon}\,{,}
\end{equation}
\begin{equation}
\alpha^{(\sigma)}\to \alpha^{(\sigma)}(\epsilon)={\cal
S}_1(\epsilon)\alpha^{(\sigma)} 
{\cal S}_1^{-1}(\epsilon)=\alpha^{(\sigma)}\cosh\epsilon + 
\alpha^{(\sigma)\dagger}\sinh\epsilon\,{,}
\end{equation}
\begin{equation}
[\beta_{q'}^{(\sigma)}, \hat{\beta}_{q'}^{(\sigma)}]=\sqrt{q'}\,
e^{-\frac{\epsilon}{2}
(\beta^{(\sigma)2}-\beta^{(\sigma)\dagger 2})}
\equiv \sqrt{q'}{\cal S}_2(\epsilon),\quad  q'\equiv e^{-\epsilon}=1/q\,{,}
\end{equation}
\begin{equation}
\beta^{(\sigma)}\to\beta^{(\sigma)}(\epsilon)=
{\cal S}_2(\epsilon)\beta^{(\sigma)}{\cal S}_2^{-1}(\epsilon)=
\beta^{(\sigma)}\cosh\epsilon
-\beta^{(\sigma)\,\dagger}\sinh\epsilon\,{,}
\end{equation}
and hermitian conjugates of Eqs. (4.3) and (4.4). For  
simplicity in Eqs. (4.2)-(4.5) we dropped the $\sigma$ index in the
$S_i(\epsilon), i=1,2$ generators. 

${\cal S}_i(\epsilon), i=1,2$ are $SU(1,1)$ group
elements. The set of operators,
$\{J_{-}^{(\sigma)}=\frac{1}{2}\alpha^{(\sigma)2},\; 
J_{+}^{(\sigma)}=\frac{1}{2}\alpha^{(\sigma)\dagger\,2}, 
\; J_{0}^{(\sigma)}=\frac{1}{2}
(\alpha^{(\sigma)\dagger}\alpha^{(\sigma)}+ \frac{1}{2})\}$ and  
$\{K_{-}^{(\sigma)}=\frac{1}{2}\beta^{(\sigma)2},\; 
K_{+}^{(\sigma)}=\frac{1}{2}\beta^{(\sigma)\dagger\,2},\; 
K_{0}^{(\sigma)}=\frac{1}{2}(\beta^{(\sigma)\dagger}\beta^{(\sigma)}+
\frac{1}{2})\}$
close the $\oplus_{(\sigma)}su(1,1)_{(\sigma)}$ algebra, indeed. 

The product of the q-deformed algebras (4.2) and (4.4) yields
\begin{equation}
\prod_{\sigma}[\alpha_q^{(\sigma)}, \hat{\alpha}_q^{(\sigma)}]
[\beta_{q'}^{(\sigma)}, \hat{\beta}_{q'}^{(\sigma)}]=
e^{\frac{\epsilon}{2}\sum_{\sigma}
[(\alpha^{(\sigma)2}
-\alpha^{(\sigma)\dagger 2})-(\beta^{(\sigma)2}-\beta^{(\sigma)\dagger
2})]} \,{.}
\end{equation}
The formal relation between thermalization theorem and q-groups 
is established by generalizing Eqs. (4.1)-(4.6) 
to infinite degrees of freedom.

The Bogoliubov transformations (4.3) and (4.5) can be
implemented for any $k$, where $k$ label the field degrees of freedom
(i.e. momentum), 
as inner automorphism for the algebra 
$su(1,1)_{(k,\sigma)}$ \cite{SOL}. To every value of the
parameter 
$\epsilon$ corresponds a copy 
$\{\alpha_k^{(\sigma)}(\epsilon), \alpha_k^{(\sigma)\,
\dagger}(\epsilon), \beta_k^{(\sigma)}(\epsilon), 
\beta_k^{(\sigma)\,\dagger}(\epsilon), \sigma=\pm;
|0(\epsilon)> \forall\,k\}$ of the original algebra 
$\{\alpha_k^{(\sigma)},
\alpha_k^{(\sigma)\,\dagger}, \beta_k^{(\sigma)}, \beta_k^{(\sigma)\,
\dagger}, \sigma=\pm; |0_M> \forall\,k\}$;  
the Bogoliubov generator can be 
thought of as the generator of the group of automorphisms of
$\oplus_{(k,\sigma)} 
su(1,1)_{(k,\sigma)}$ parameterized by $\epsilon$. 
In this way, in the limit of infinite degrees of freedom, 
relations (4.1) and (4.6) become (for a n-dimensional space) 
\begin{equation}
[\alpha_k^{(\sigma)}, \alpha_{k'}^{(\sigma')\,\dagger}]=
[\beta_k^{(\sigma)}, \beta_{k'}^{(\sigma')\,
\dagger}]=\delta_{\sigma\sigma'} 
\,\delta(k-k')\,{,} \quad \sigma, \sigma'=\pm\,{,}
\end{equation}
$$
\prod_{k,\sigma}[\alpha_{k,q}^{(\sigma)}, \hat{\alpha}_{k,q}^{(\sigma)}]
[\beta_{k,q'}^{(\sigma)}, \hat{\beta}_{k,q'}^{(\sigma)}]\to
\exp\left[\frac{1}{2}\frac{V}{(2\pi)^{n-1}}\sum_{\sigma}\right.$$
\begin{equation}
\left.
\int d^{n-1}p\;\epsilon(p)
[(\alpha_p^{(\sigma)2}-\alpha_p^{(
\sigma)\dagger 2})-(\beta_p^{(\sigma)2}-\beta_p^{(\sigma)\dagger 2})]
\right]\equiv G(\epsilon)\,{.}
\end{equation}
Note that for simplicity $q$ 
and $q'$ denote $q_k$ and $q_k'$. 

In order to write the Bogoliubov generator $G(\epsilon)$ in (4.8) in
a more 
convenient form and to establish the connection with the results of
Section 2,  
let us write it in terms of the following independent
operators $d_k^{(\sigma)}, \bar{d}_{\tilde k}^{(-\sigma)}$, related to
the  
operators
$\alpha_k^{(\sigma)}, \beta_k^{(\sigma)}$:
\begin{equation}
\alpha_k^{(\sigma)}=
\frac{1}{\sqrt{2}}(d_k^{(\sigma)}+\bar{d}_{\tilde k}^{(-\sigma)}),
\quad \beta_k^{(\sigma)}=
\frac{1}{\sqrt{2}}(d_k^{(\sigma)}-\bar{d}_{\tilde k}^{(-\sigma)})\,{.}
\end{equation}
We recall that $d_k^{(\sigma)}$ and $\bar{d}_{\tilde k}^{(-\sigma)}$ 
are linear combinations of the Minkowski annihilation operators alone 
(cf. Eq. (2.11)),
annihilate the Minkowski vacuum $|0_M>$ (cf. Eq. (2.12)) 
and satisfy the canonical commutation relations. Moreover
$k=(\tilde{\omega}, \vec{k})$ and $\tilde{k}=(\tilde{\omega},
-\vec{k})$. 
In terms of them, the Bogoliubov generator (4.8) reads
\begin{equation}
G(\epsilon)=\exp\left[\frac{V}{(2\pi)^{(n-1)}}\sum_{\sigma}\int d^{n-1}p\;
\epsilon(p)
[d_p^{(\sigma)}\bar{d}_{\tilde p}^{(-\sigma)}
-d_p^{(\sigma)\,\dagger}\bar{d}_{\tilde p}^{(-\sigma)\,\dagger}]\right]\,{,}
\end{equation}
where $d^{n-1}p=d\tilde{\omega}d\vec{p}$, $p=(\tilde{\omega}, \vec{p})$ and
$\tilde{p}=(\tilde{\omega},-\vec{p})$. $G(\epsilon)$
is an unitary operator:
$G^{-1}(\epsilon)=G(-\epsilon)=G^{\dagger}(\epsilon)$ 
and induces the following Bogoliubov transformations
\begin{equation}
d_k^{(\sigma)}\to
d^{(\sigma)}_k(\epsilon)=G(\epsilon)d_k^{(\sigma)}G^{-1} 
(\epsilon)\,=\,d_k^{(\sigma)}\cosh\epsilon(k)\,+
\,\bar{d}_{\tilde k}^{(-\sigma)\,\dagger}
\sinh\epsilon(k)\,{,}
\end{equation}
\begin{equation}
\bar{d}_{\tilde k}^{(-\sigma)\dagger}\to 
\bar{d}_{\tilde k}^{(-\sigma)\dagger}(\epsilon)=
G(\epsilon)\bar{d}_{\tilde k}^{(-\sigma)\dagger}G^{-1}
(\epsilon)\,=\,d_k^{(\sigma)}\sinh\epsilon(k)\,+
\,\bar{d}_{\tilde k}^{(-\sigma)\,\dagger}
\cosh\epsilon(k)\,{.}
\end{equation}
$d_k^{(\sigma)}(\epsilon)$ and $\bar{d}_{\tilde
k}^{(-\sigma)}(\epsilon)$  
satisfy for any $\epsilon$ 
the canonical commutation relations, i.e. satisfy the same
algebra 
of the operators $d_k^{(\sigma)}$ and $\bar{d}_{\tilde
k}^{(-\sigma)}$. 

Transformations (4.11) and (4.12) are recognized to be the
transformations  
(2.9) and (2.10) provided
\begin{equation}
d_k^{(\sigma)}(\epsilon)\equiv b_k^{(\sigma)},\quad 
\bar{d}_{\tilde k}^{(-\sigma)}(\epsilon)\equiv\bar{b}_{\tilde
k}^{(-\sigma)}\,{,} 
\end{equation}
\begin{equation}
\sinh\epsilon(k)\,=
\,\left(\frac{1}{e^{2\pi\tilde{\omega}/a}-1}\right)^{1/2}\,{,}
\end{equation}
with $k=(\tilde{\omega}, \vec{k})$.
Eqs. (4.13) and (4.14) are the wanted result: they express the relation
between the deformation parameter $q_k=e^{\epsilon(k)}$ and the
accelerated frame operators 
$b_k^{(\sigma)}$ and $\bar{b}_{\tilde k}^{(-\sigma)}$ (Eq. (4.13)),
and the coefficient of the Bogoliubov transformations (Eq. (4.14))
relating inertial and accelerated frame operators (cfr. Eqs.
(2.9) and (2.10)). Moreover, Eq. (4.14) shows that
the Rindler acceleration is related to the deformation parameter
so that we can write now $q_{k, a}=e^{\epsilon(\tilde{\omega}, a)}$, and
the link between thermalization theorem and q-WH algebra is thus established.
Note that since, for a given $\tilde{\omega}$, $q_{k,a}\to 1$ for
$a\to 0$, the Rindler acceleration $a$ plays in fact the role
of deformation parameter. 

In conclusion, our results may be summarized as follows.
 
The Hilbert space $\cal  H$ of the basis vectors associated to the
Minkowski  
space (inertial frame) is build by repeated action 
of $(d_k^{(\sigma)\,\dagger}, \bar{d}_{\tilde k}^{(-\sigma)\,\dagger})$ 
on the vacuum state
$|0_M>$. Bogoliubov transformations,
Eqs. (4.11) and (4.12), relate vectors of $\cal  H$ to vectors of
another 
Hilbert space labeled by $\epsilon=\epsilon(\tilde{\omega},a)$, 
${\cal H}_{\epsilon}$,
for a fixed value of the acceleration $a$. The relation
between these spaces is established by the generator $G(\epsilon)$
that 
maps $\cal  H$ in ${\cal H}_{\epsilon}$, $G(\epsilon): {\cal H} \to 
{\cal H}_{\epsilon}$ (for fixed $a$). In particular, for the vacuum 
state $|0_M>$ one has
\begin{equation}
|0(\epsilon)>\,=\,G(\epsilon)\,|0_M>\,{,}
\end{equation}
where $|0(\epsilon)>$ is the vacuum state of the Hilbert space 
${\cal H}_{\epsilon}$ associated to the accelerated frame. In other 
words, $|0(\epsilon)>\equiv |0_R>$. 

By inverting Eq. (4.15) and using the Gaussian decomposition \cite{PER}, 
the vacuum state of the inertial frame can be written as
\begin{equation}
|0_M>=Z\,\exp\left[{\sum_{\sigma} \int d \tilde{\omega}
\int d^{n-2}p\;\tanh\epsilon(\tilde{\omega},a)b_p^{(\sigma)\dagger}
\bar{b}_{\tilde p}^{(-\sigma)\dagger}}\right]\,|0_R>\,{,}
\end{equation}
where $\tanh\epsilon(\tilde{\omega},a)=e^{-\pi\tilde{\omega}/a}$ and
$Z=\exp[{\int d\tilde{\omega}\int
d^{n-2}p\;\ln\cosh\epsilon(\tilde{\omega},a)}]$. 
As pointed out in Section 2, the relation between the two vacua, $|0_M>$ 
and 
$|0_R>$, follows by the ansatz $|0_M>=F(b, \bar{b})|0_R>$. In the
q-groups  
approach such a relation, Eq. (4.16), is directly established by the 
Bogoliubov generator, i.e. by Eq. (4.15).

The number of modes of type $b_k^{(\sigma)}$ is given, for each fixed 
value of $a$, by
\begin{equation}
<0_M|b_k^{(\sigma)\dagger}b_k^{(\sigma)}|0_M>=
\sinh^2\epsilon(\tilde{\omega},a)=\frac{1}{e^{2\pi\tilde{\omega}/a} -1}
\end{equation}
and similarly for the modes of type $\bar{b}_{\tilde k}^{(-\sigma)}$.
Moreover, $<0(\epsilon)|0(\epsilon)>=1, \forall\epsilon$. We note  
that in the infinite-volume limit, we have 
\begin{equation}
<0(\epsilon)|0_M>\to 0\quad as\,\,V\to\infty, \forall\epsilon\,{,}
\end{equation}
\begin{equation}
<0(\epsilon)|0(\epsilon^{\prime})>\to 0 \quad as\,\,V\to\infty, 
\forall\epsilon,\epsilon^{\prime}, \epsilon\ne\epsilon^{\prime}\,{,}
\end{equation}
i.e., the Hilbert spaces ${\cal  H}$ and ${\cal H}_{\epsilon}$
become orthogonal in the infinite volume limit. In this limit, as $\epsilon$ 
evolves by varying the Rindler acceleration $a$,
one runs over a variety of infinitely many unitarily inequivalent
representations of the canonical commutation relations. $\epsilon^{\prime}$
in Eq. (4.19) corresponds to
the Rindler acceleration $a^{\prime}$.
 
Because the quantum deformation parameter acts as a label for the unitarily
inequivalent representations in QFT \cite{DIS}, the mapping between different 
(i.e. labeled by different values of q) representations
being performed by the Bogoliubov transformations, at finite $V$ 
the Rindler accelerations may be taken as a label for such unitarily
inequivalent representations, as well.

\section{Conclusions}

In this paper, n-dimensional QFT developed for accelerated
coordinates in flat space-time and its thermal properties have been
analyzed 
in terms of q-groups. We have shown that the Bogoliubov
generator relating 
the annihilation 
and creation operators in the accelerated frame  
to the ones in the inertial frame, can be expressed in terms of
commutators of the q-WH algebra.  Then, quantum deformation
parameter turns out to related to the Rindler acceleration, which
acts as a label for the unitarily inequivalent representations
of the canonical commutation relations. The link between thermalization
theorem and q-WH algebra thus emerges.

A possible application of these results to different 
space-time, for instance to Schwarzschild space-time,  
is certainly of interest. In this case, the procedure of quantization
of a free scalar field follows the same prescription of the one 
in the Rindler 
manifold and the surface of gravity characterizing the black hole
plays the role of the Rindler acceleration. Then, one can conclude 
that the deformation parameter
is related to the surface of gravity.

As our analysis shows, quantum deformations are mathematical structures
underlying the QFT in "curved" space-time. This strongly suggests a deep
connection between q-groups and quantum gravity, and encouraging results
in this direction have been obtained in different contexts \cite{KEM, MAG}. 
However, much
work is still needed to obtain a full understanding of
this subject. 

\bigskip

{\bf Acknowledgments}

The author thanks G. Scarpetta and G. Vitiello for fruitful discussions.
This work has been supported by Ministero dell'Universit\'a e della Ricerca 
Scientifica.

\end{document}